\documentclass[12pt]{article}
\usepackage{graphicx,graphics}
\usepackage{amsmath}
\usepackage{amsfonts}
\usepackage{amssymb}

\begin{document}

\title{A practical approach to solve coupled systems of nonlinear partial
differential equations.}
\author{Alvaro Salas\thanks{Universidad de
Caldas, Department of Mathematics, Universidad Nacional de Colombia,
Manizales. \emph{email} : ahsalass2@unal.edu.co}\\
Cesar G\'omez \thanks{Department of Mathematics, Universidad
Nacional de Colombia, sede Bogot\'a. \emph{email} :
cagomezsi@unal.edu.co}}
\date{}
\maketitle

\begin{abstract}
In this paper we present the tanh method to obtain exact solutions to
coupled MkDV system. This method may be applied to a variety of coupled
systems of nonlinear ordinary and partial differential equations.
\end{abstract}


\section{Introduction}

The search of exact solutions to coupled systems of nonlinear partial
differential equations is of great importance, because these systems appear
in complex physics phenomena, mechanics, chemistry, biology and engineering.
A variety of powerful and direct methods have been developed in this
direction. The principal objective of this paper, is to present the
application of tanh method in solving coupled systems of two equations. The
tanh method, developed by Malfliet in [2], proved to be effective and
reliable for several nonlinear problems.

This method works effectively if the equation involves sine, cosine,
hyperbolic sine, hyperbolic cosine functions, and exponential functions.

\section{The tanh method}

Consider a system of two coupled PDE's in the variables $t$, $x$

\begin{equation}
\left\{ {
\begin{array}{l}
P(u,v,u_{x},v_{x},u_{t},v_{t},u_{xt},v_{xt},u_{xx},v_{xx},\ldots )=0 \\
Q(u,v,u_{x},v_{x},u_{t},v_{t},u_{xt},v_{xt},u_{xx},v_{xx},\ldots
)=0\,.
\end{array}
}\right. \,\,\,  \label{eq1}
\end{equation}
Using the wave transformation
\begin{equation}
\begin{array}{l}
u(x,t)=u(\xi ),\,\,v(x,t)=v(\xi ), \\
\quad \,\,\,\,\xi =x+\lambda t,
\end{array}
\label{eq2}
\end{equation}
\noindent where$\lambda $ ia a constant, system (\ref{eq1}) reduces
to a system of two ordinary nonlinear differential equations
\begin{equation}
\left\{ {
\begin{array}{l}
p(u(\xi ),u^{\prime }(\xi ),v^{\prime \prime }(\xi ),\ldots )=0 \\
q(v(\xi ),v^{\prime }(\xi ),v^{\prime \prime }(\xi ),\ldots )=0.
\end{array}
}\right.  \label{eq3}
\end{equation}
\noindent The tanh method [2] is based on the idea of looking for
solutions to system ( \ref{eq3}) in the form
\begin{equation}  \label{eq4}
u(\xi ) = \sum\limits_{i = 0}^m {a_i \varphi ^i(\xi )} ,\,\,\,v(\xi
) = \sum\limits_{j = 0}^n {b_j \varphi ^j(\xi )},
\end{equation}
\noindent where the new variable $\varphi = \varphi (\xi )$
satisfies the Riccati equation
\begin{equation}  \label{eq5}
\varphi ^{\prime}= \varphi ^2 + k\,
\end{equation}
\noindent whose solutions are given by
\begin{equation}
\varphi (\xi )=\left\{
\begin{array}{ll}
\,\,\,\,-1/\xi & \ k=0 \\
\,\,\sqrt{k}\tan (\sqrt{k}\xi ) & \ k>0 \\
-\sqrt{k}\cot (\sqrt{k}\xi )\quad & k>0 \\
-\sqrt{-k}\tanh (\sqrt{-k}\xi )\,\,\, & k<0 \\
-\sqrt{-k}\mbox{coth}(\sqrt{-k}\xi ) & k<0.
\end{array}
\right.  \label{eq6}
\end{equation}
The integers $m$ and $n$ can be determined balancing the highest
derivative terms with nonlinear terms in (\ref{eq3}). Substituting
(\ref{eq4}) along with (\ref{eq5}) into (\ref{eq3}) and collecting
all terms with the same power $\varphi ^i$, we get two polynomials
in the variable $\varphi $. Equating the coefficients of these
polynomials to zero, we obtain a system of algebraic equations, from
which the constants $a_i $, $b_j $, $\mu $, $ \lambda $, $k$ are
obtained explicitly. This allows us to obtain solutions to system
(\ref{eq1}).

\section{Exact solutions for coupled MkdV equations}

The system of coupled MkdV equations [3] reads
\begin{equation}
\left\{ {
\begin{array}{l}
u_{t}=\frac{1}{2}u_{xxx}-3u^{2}u_{x}+\frac{3}{2}v_{xx}+3(uv)_{x}-3\eta u_{x}
\\
v_{t}=-v_{xxx}-3vv_{x}-3u_{x}v_{x}+3u^{2}v_{x}+3\eta v_{x}\,,
\end{array}
}\right. \,\,  \label{eq7}
\end{equation}
\noindent where $\eta $ is a constant. We apply the transformation
\begin{equation}  \label{eq8}
u(x,t) = u(\xi ),\,\,v(x,t) = v(\xi ),\,\,\xi = x + \lambda
t.\,\,\,\,
\end{equation}
After substitution of (\ref{eq8}) into (\ref{eq7}) we get the
following system of nonlinear ode's :
\begin{equation}
\left\{ {
\begin{array}{l}
2(3\eta +\lambda )u^{\prime }+6u^{2}u^{\prime }-6u^{\prime }v-6uv^{\prime
}-3v^{\prime \prime }-u^{\prime \prime \prime }=0 \\
(\lambda -3\eta )v^{\prime }-3u^{2}v^{\prime }+3vv^{\prime
}+3u^{\prime }v^{\prime }+v^{\prime \prime \prime }=0.
\end{array}
}\right. \,\,\,  \label{eq9}
\end{equation}
We now substitute (\ref{eq4}) into (\ref{eq9}) and using (\ref{eq5})
we obtain a system in the variable $\varphi $ :
\begin{equation}
\left\{ {
\begin{array}{l}
c_{1}\varphi ^{m+n+1}+c_{2}\varphi ^{3m+1}+c_{3}\varphi ^{m+3}+ \\
c_{4}\varphi ^{m+2}+c_{5}\varphi ^{m+1}+c_{6}=0. \\
d_{1}\varphi ^{2m+n+1}+d_{2}\varphi ^{m+n+2}+d_{3}\varphi ^{2n+1}+ \\
d_{4}\varphi ^{n+3}+d_{5}\varphi ^{n+1}+d_{6}=0.
\end{array}
}\right.  \label{eq10}
\end{equation}
Balancing the highest order terms in (\ref{eq9}) gives
\begin{equation*}
\left\{ {
\begin{array}{l}
3m+1=m+n+1 \\
2m+n+1=m+n+2
\end{array}
}\right.
\end{equation*}
\noindent and then $m = 1$ and $n = 2$. Therefore we seek solutions
to (\ref {eq8}) in the form
\begin{equation}
\begin{array}{l}
u(\xi )=a_{0}+a_{1}\varphi (\xi ),\, \\
v(\xi )=b_{0}+b_{1}\varphi (\xi )+b_{2}\varphi ^{2}(\xi )\,.
\end{array}
\label{eq11}
\end{equation}
Substituting (\ref{eq10}) into (\ref{eq9}), and using (\ref{eq5}) we
obtain an algebraic system in the variables $a_0 $, $a_1 $, $b_0 $,
$b_1 $, $b_2 $, $k$ and $\lambda $. Solving it with the aid of
\textit{Mathematica} we get the following solutions :
\begin{itemize}
\item{First Family} : For $\lambda =k+3a_{0}^{2}$, $~a_{-1}=-1$,~\  $
a_{0}=a_{0}$, $\ b_{0}=\eta $, $\ b_{1}=-2a_{0}$ and $b_{2}=0$ :
\begin{eqnarray*}
&&\left\{ {
\begin{array}{l}
u_{1}=a_{0}+\sqrt{-k}\tanh (\sqrt{-k}(x+(k+3a_{0}^{2})t) \\
v_{1}=\eta +2a_{0}\sqrt{-k}\tanh (\sqrt{-k}(x+(k+3a_{0}^{2})t))
\end{array}
}\right.  \\
&&\left\{ {
\begin{array}{l}
u_{2}=a_{0}+\sqrt{-k}\coth (\sqrt{-k}(x+(k+3a_{0}^{2})t) \\
v_{2}=\eta +2a_{0}\sqrt{-k}\coth (\sqrt{-k}(x+(k+3a_{0}^{2})t))
\end{array}
}\right.  \\
&&\left\{ {
\begin{array}{l}
u_{3}=a_{0}-\sqrt{k}\tan (\sqrt{k}(x+(k+3a_{0}^{2})t) \\
v_{3}=\eta -2a_{0}\sqrt{k}\tan (\sqrt{k}(x+(k+3a_{0}^{2})t))
\end{array}
}\right.  \\
&&\left\{ {
\begin{array}{l}
u_{4}=a_{0}-\sqrt{k}\cot (\sqrt{k}(x+(k+3a_{0}^{2})t) \\
v_{4}=\eta -2a_{0}\sqrt{k}\cot (\sqrt{k}(x+(k+3a_{0}^{2})t))
\end{array}
}\right.
\end{eqnarray*}
\item{Second Family} : For $\lambda = - 7k$, $a_0 = 0$, $a_1 = 3$,
$b_0 = - \frac{10k}{3}$, $b_1 = 0$ and $b_2 = 2$ :
\begin{equation*}
\left\{ {
\begin{array}{l}
u_{5}=3\sqrt{-k}\tanh (\sqrt{-k}(x-7kt)) \\
v_{5}=\eta -\frac{10k}{3}-2k\tanh ^{2}(\sqrt{-k}(x-7kt))
\end{array}
}\right.
\end{equation*}
\begin{equation*}
\left\{ {
\begin{array}{l}
u_{6}=3\sqrt{-k}\coth (\sqrt{-k}(x-7kt)) \\
v_{6}=\eta -\frac{10k}{3}-2k\coth ^{2}(\sqrt{-k}(x-7kt))
\end{array}
}\right.
\end{equation*}
\begin{equation*}
\left\{ {
\begin{array}{l}
u_{7}=3\sqrt{k}\tan (\sqrt{k}(x-7kt)) \\
v_{7}=\eta -\frac{10k}{3}+2k\tan ^{2}(\sqrt{k}(x-7kt))
\end{array}
}\right.
\end{equation*}
\begin{equation*}
\left\{ {
\begin{array}{l}
u_{8}=3\sqrt{k}\cot (\sqrt{k}(x-7kt)) \\
v_{8}=\eta -\frac{10k}{3}+2k\cot ^{2}(\sqrt{k}(x-7kt))
\end{array}
}\right.
\end{equation*}
\item{Third Family} : $\lambda =-2k$, $a_{0}=0$, $\ a_{1}=-2$, $
~b_{0}=\eta $,$~~b_{1}=0$,$~~b_{2}=2\quad $:
\begin{equation*}
\left\{ {
\begin{array}{l}
u_{9}=2\sqrt{-k}\tanh (\sqrt{-k}(x-2kt) \\
v_{9}=\eta -2k\tanh ^{2}(\sqrt{-k}(x-2kt))
\end{array}
}\right.
\end{equation*}
\begin{equation*}
\left\{ {
\begin{array}{l}
u_{10}=2\sqrt{-k}\,\coth (\sqrt{-k}(x-2kt) \\
v_{10}=\eta -2k\coth ^{2}(\sqrt{-k}(x-2kt))
\end{array}
}\right.
\end{equation*}
\begin{equation*}
\left\{ {
\begin{array}{l}
u_{11}=-2\sqrt{k}\,\tan (\sqrt{k}(x-2kt) \\
v_{11}=\eta +2k\tan ^{2}(\sqrt{k}(x-2kt))
\end{array}
}\right.
\end{equation*}
\begin{equation*}
\left\{ {
\begin{array}{l}
u_{12}=-2\sqrt{k}\,\cot (\sqrt{k}(x-2kt) \\
v_{12}=\eta +2k\cot ^{2}(\sqrt{k}(x-2kt))
\end{array}
}\right.
\end{equation*}
\item{Fourth Family} : $\lambda = k$, $a_0 = 0$, $a_1 = - 1$, $b_0 =
\eta - 2k$, $b_1 = 0$, $b_2 = - 2$ :
\begin{equation*}
\left\{ {
\begin{array}{l}
u_{13}=\sqrt{-k}\tanh (\sqrt{-k}(x+kt) \\
v_{13}=\eta -2k+2k\tanh ^{2}(\sqrt{-k}(x+kt))
\end{array}
}\right.
\end{equation*}
\begin{equation*}
\left\{ {
\begin{array}{l}
u_{14}=\sqrt{-k}\,\mbox{coth}(\sqrt{-k}(x+t) \\
v_{14}=\eta -2k+2k\coth ^{2}(\sqrt{-k}(x+kt))
\end{array}
}\right.
\end{equation*}
\begin{equation*}
\left\{ {
\begin{array}{l}
u_{15}=-\sqrt{k}\,\tan (\sqrt{k}(x+kt) \\
v_{15}=\eta -2k-2k\tan ^{2}(\sqrt{k}(x+kt))
\end{array}
}\right.
\end{equation*}
\begin{equation*}
\left\{ {
\begin{array}{l}
u_{16}=-\sqrt{k}\,\cot (\sqrt{k}(x+kt) \\
v_{16}=\eta -2k-2k\cot ^{2}(\sqrt{k}(x+kt))
\end{array}
}\right.
\end{equation*}
\end{itemize}

\section{Conclusions}

By using the tanh method [2], we obtained sixteen solutions to the coupled
MkdV equations. The method is applicable to other coupled systems. There are
other methods to solve nonlinear differential equations, for example the
tanh-coth method [1] and the projective Riccati equation method [4][5].

\end{document}